# Characterization and valuation of uncertainty of calibrated parameters in stochastic decision models

Fernando Alarid-Escudero[1], Amy B. Knudsen[2], Jonathan Ozik[3,4], Nicholson Collier[3,4], Karen M. Kuntz[5]

[1] Drug Policy Program, Center for Research and Teaching in Economics (CIDE) – CONACyT, Aguascalientes, Aguascalientes, Mexico, 20313
[2] Institute for Technology Assessment, Massachusetts General Hospital, Harvard Medical School, Boston, MA, 02114
[3] Decision and Infrastructure Sciences Division, Argonne National Laboratory, Argonne, IL, 60439
[4] Consortium for Advanced Science and Engineering, University of Chicago, Chicago, IL, 60637
[5] Division of Health Policy and Management, University of Minnesota School of Public Health, Minneapolis, MN, 55455

**Corresponding author:** Fernando Alarid-Escudero, PhD, fernando.alarid@cide.edu, Drug Policy Program, Center for Research and Teaching in Economics (CIDE) – CONACyT, Aguascalientes, Aguascalientes, Mexico

## Abstract

**Background.** We evaluated the implications of different approaches to characterize uncertainty of calibrated parameters of stochastic decision models (DMs) and quantified the value of such uncertainty in decision making. **Methods**. We used a microsimulation DM of colorectal cancer (CRC) screening to conduct a cost-effectiveness analysis (CEA) of a 10-year colonoscopy screening. We calibrated the natural history (NH) model of CRC to epidemiological data with different degrees of uncertainty and obtained the joint posterior distribution of the parameters using a Bayesian approach. We conducted a probabilistic sensitivity analysis (PSA) on all the model parameters with different characterizations of uncertainty of the calibrated parameters and estimated the value of uncertainty of the different characterizations with a value of information analysis. All analyses were conducted using high performance computing (HPC) resources running the Extreme-scale Model Exploration with Swift (EMEWS) framework. **Results.** The posterior distribution had high correlation among some parameters. The parameters of the Weibull hazard function for the age of onset of adenomas had the highest posterior correlation of -0.958. Considering full posterior distributions and the maximum-a-posteriori estimate of the calibrated parameters, there is little difference on the spread of the distribution of the CEA outcomes with a similar expected value of perfect information (EVPI) of $653 and $685, respectively, at a WTP of $66,000/QALY. Ignoring correlation on the posterior distribution of the calibrated parameters, produced the widest distribution of CEA outcomes and the highest EVPI of $809 at the same WTP. **Conclusions.** Different characterizations of uncertainty of calibrated parameters have implications on the expect value of reducing uncertainty on the CEA. Ignoring inherent correlation among calibrated parameters on a PSA overestimates the value of uncertainty.

*Keywords:* characterization of uncertainty; calibration; Bayesian; value of information analysis; decision-analytic models; high-performance computing; EMEWS

## BACKGROUND

Decision models (DMs) are commonly used in cost-effectiveness analysis where uncertainty in the parameters is inherent (1). The impact of parameter uncertainty can be assessed with probabilistic sensitivity analysis (PSA) to characterize decision uncertainty (i.e., the probability of a strategy being cost-effective) and to quantify the value of potential future research by determining the potential consequences of a decision with value of information (VOI) analysis (2,3).

The parameters of DMs can be split into two categories, those that are obtained from the literature or can be estimated from available data (i.e., external parameters), and those that need to be estimated through calibration (i.e., calibrated parameters). Uncertainty of external parameters is estimated either from individual-level or aggregated data that directly inform the parameters of interest. There are recommendations on the type of distributions that characterize their uncertainty based on the characteristics of the parameters or the statistical model used to estimate them (3). For example, a probability could be modeled with a beta distribution and a relative risk with a lognormal distribution (4). For calibrated parameters, no such data exist that can directly inform their uncertainty because a research study hasn't been conducted or is unfeasible to conduct, or because the parameters reflect unobservable phenomena, as is often the case in natural history models of chronic diseases (5–8) or in infectious disease dynamic models (9). The choice of distribution for these parameters is often less clear. One option is to define uniform distributions with wide bounds or generate informed distributions based on moments of the calibrated parameters, such as the mean and standard error. However, the impact of these approaches to characterize uncertainty of calibrated parameters on decision uncertainty and VOI has not been studied.

Model calibration is the process of estimating these parameters by matching model outputs to observed clinical or epidemiological data (known as calibration targets) (1,10–12). While there are several approaches for searching the parameter space in the calibration process, most approaches are insufficient when one wants to characterize the uncertainty in the calibrated model parameters because they do not provide interval estimates. For example, direct-search optimization algorithms like Newton-Raphson, Nelder-Mead (13), simulated annealing or genetic algorithms (14) treat the calibration targets as if they were known with certainty, so are primarily useful when one wants to identify a single or a set of parameter sets that yield good fit to the targets (10).

A sample of calibrated parameter sets that correctly characterizes the uncertainty of all of the calibration target data has to be obtained from their joint distribution conditional on the calibrated targets. To obtain this joint distribution of the calibrated parameters, the calibration could be seen as a statistical estimation problem under two different frameworks, through maximum likelihood (ML) or Bayesian methods. ML can fail in obtaining interval estimates for stochastic models and when the calibration problem is non-identifiable (12,15); thus, we will focus on a Bayesian approach.

In a Bayesian-calibration framework, instead of obtaining a single best-fitting parameter or a set of good parameter sets, the solution is a joint (posterior) probability distribution (10,16,17) of calibrated parameters that adequately characterizes the uncertainty of all of the calibration targets (10,18–21) and prior information about the model parameters (22). A Bayesian approach is able to characterize uncertainty of calibrated parameters even in the presence of non-identifiability (12, 15,23). Despite their suitability to correctly characterize uncertainty of calibrated model parameters, Bayesian methods are computationally expensive because of the need to evaluate the model thousands and sometimes millions of times. The computational burden of Bayesian methods does not seem to be an impediment when calibrating deterministic DM (17,24) but they become more difficult to apply to stochastic DMs, such as microsimulation, discrete-event simulation and agent-based models, which has limited its use to a few of them (5).

However, the increasing availability of high-performance computing (HPC) systems both in academic and commercial settings expands the use of such systems for model calibration and model exploration of stochastic DMs to a broader audience. The Extreme-scale Model Exploration with Swift (EMEWS) framework facilitates such large-scale model calibration and exploration on HPC resources (25). EMEWS (emews.org) offers the capability to run very large, highly concurrent ensembles of stochastic DMs of varying types while supporting a wide class of calibration algorithms, including those increasingly available to the community via Python and R libraries, on high-performance computing clusters.

The purpose of our study is twofold. First, to use recently developed HPC capabilities to characterize uncertainty of calibrated parameters of a microsimulation model of the natural history (NH) of colorectal cancer (CRC). Second, to explore the impact of different approaches to characterizing the uncertainty of calibrated parameters on decision uncertainty and to quantify the value of eliminating parameter uncertainty using VOI analysis using a stochastic DM for the cost-effectiveness of CRC screening strategies.

**METHODS**

We developed a microsimulation model of the NH of CRC and calibrated it using a Bayesian approach. Instead of using the posterior means to represent the best estimates of each calibrated parameter, we obtained the posterior distribution using a Bayesian approach that represents the joint uncertainty of all of the calibrated parameters that can then be used in a PSA. We then overlaid a simple CRC screening strategy onto the NH model and conducted a cost-effectiveness analysis (CEA) of the CRC screening strategy, including a PSA. We then evaluated the impact of different approaches to characterize uncertainty of calibrated parameters on the joint distribution of incremental costs and incremental effects of the screening strategy compared with no screening through a PSA while also accounting for the uncertainty of the external parameters (e.g., test characteristics, costs, etc.). We also quantified the value of eliminating all parameter uncertainty by calculating the expected value of perfect information (EVPI).

*Microsimulation model of the natural history of CRC*

We developed a state-transition microsimulation model of the NH of CRC implemented in R (26) based on a structure of a previously implemented deterministic model (12). The progression between health states follows a continuous-time age-dependent Markov process. There are two age-dependent transition intensities (i.e., transition rates), $\lambda_1(a)$ and $\mu(a)$, that govern the age of onset of adenomas and non-cancer-specific mortality, respectively. Following Wu et al. (27), we specify $\lambda_1(a)$ as a Weibull hazard with the following specification

$$\lambda_1(a) = l\gamma a^{\gamma-1},$$

where $l$ and $\gamma$ are the scale and shape parameters of the Weibull hazard function, respectively. The model simulates two adenoma categories: small (adenoma smaller than 1 cm in size) and large (adenoma larger than or equal to 1 cm in size). All adenomas start small and can transition to the large size category at a constant annual rate $\lambda_2$. Large adenomas may become preclinical CRC at a constant annual rate $\lambda_3$. Both, small and large adenomas may progress to preclinical CRC, although most will not in an individual's lifetime. Early preclinical cancers (preclinical stages I and II) progress to late stages (preclinical stages III and IV) at a constant annual rate $\lambda_4$ and could become symptomatic at a constant annual rate $\lambda_5$. Late preclinical cancer could become symptomatic at a constant annual rate $\lambda_6$. After clinical detection, the model simulates the survival time to death from early and late CRC using time-homogeneous mortality rates, $\lambda_7$ and $\lambda_8$, respectively. In total, the model has nine health states: normal, small adenoma, large adenoma, preclinical early CRC, preclinical late CRC, clinical early CRC, clinical late CRC, CRC death and death from other causes. The state-transition diagram of the model is shown in Figure 1.

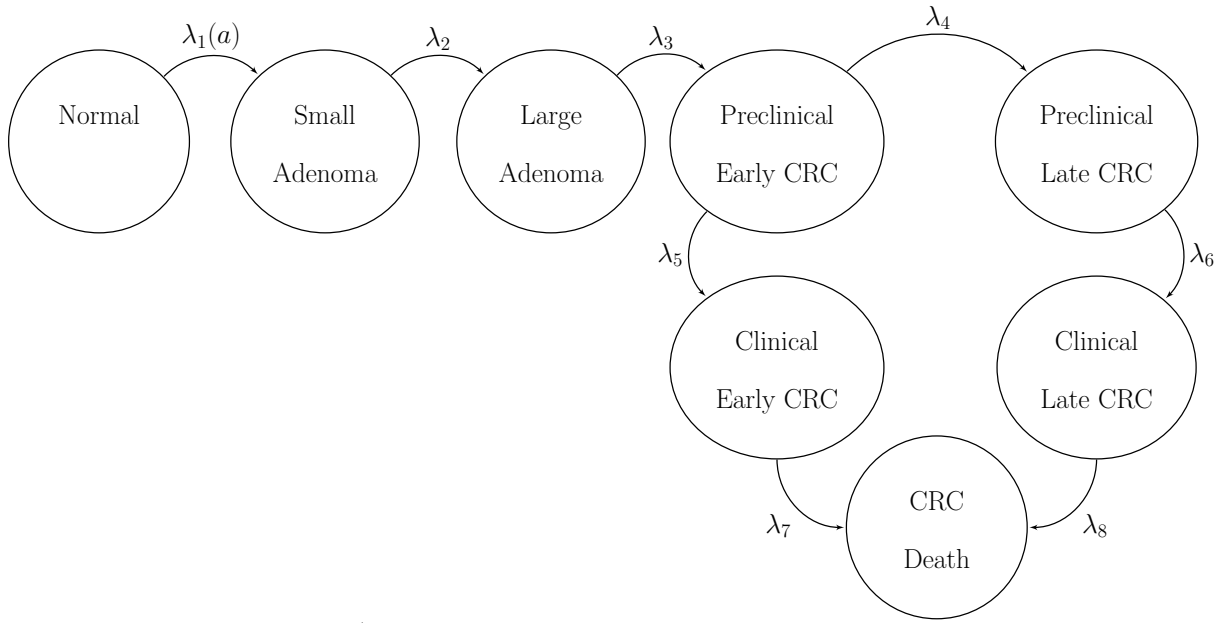

Figure 1. State-transition diagram of the nine-state microsimulation model of the natural history of colorectal cancer. Individuals in all health states face an age-specific mortality of dying from other causes (state not shown).

The continuous-time age-dependent Markov process of this NH model of CRC can be represented by an age-dependent $9 \times 9$ transition intensity matrix, $Q(a)$. To translate $Q(a)$ to discrete time, we compute the annual-cycle age-dependent transition probability matrix, $P(a, t)$, using the Kolmogorov differential equations (7,28,29)

$$P(a, t) = \text{Exp}(tG(a)),$$

where $t = 1$ and $\text{Exp}()$ is the matrix exponential. In discrete time, the microsimulation model of the NH of CRC allows transitions across multiple health states in a single year. That is, small and large adenomas may progress to preclinical or clinical CRC and preclinical cancers may progress through early and late stages.

We simulated a hypothetical cohort of 50-year-old women in the US over a lifetime. The cohort starts the simulation with a prevalence of adenoma of $p_{adeno}$, from which a proportion, $p_{small}$, correspond to small adenomas and prevalence of preclinical early and late CRC of 0.12% and 0.08%, respectively. The simulated cohort is at risk of non-CRC mortality from all health states. We obtained all-cause mortality, $\mu(a)$, from 2014 US life tables.(30)

*Calibration targets*

To calibrate the parameters of the microsimulation model of the NH of CRC, we conducted a confirmatory simulation by choosing a set of parameters based on plausible estimates from Wu et al. (27) and using the model to simulate four different age-specific targets, including adenoma prevalence, proportion of small adenomas and CRC incidence for early and late stages, which resemble commonly used calibration targets for this type of model (5, 24,31,32). To simulate the calibration targets, we ran the microsimulation model 100 times to get a stable estimate of the standard errors (SEs) using the fixed values in Table 1. We then aggregated the results across all 100 outputs to compute their mean and SE. Different calibration targets could have different level of uncertainty given the amount of data to compute their summary measures. Accordingly, we simulated different types of targets based on cohorts of different sizes to account for different variations in the amount of data on different calibration targets. Adenoma-related targets were based on a cohort of 500 individuals and cancer incidence targets were based on a cohort of 100,000 individuals.

| Symbol | Description | Value | Source | Prior Distribution | Calibrated |
|---|---|---|---|---|---|
| **Initial state of 50-year-old cohort** | | | | | |
| | Proportions | | | | |
| $p_{adeno}$ | Prevalence of adenoma at age 50 | 0.25 | (33) | Beta(3, 8) | Yes |
| $p_{small}$ | Proportion adenomas that are small at age 50 | 0.71 | (27) | Beta(6, 3) | Yes |
| - | Prevalence of preclinical early CRC at age 50 | 0.12 | | Fixed | No |
| - | Prevalence of preclinical late CRC at age 50 | 0.08 | | Fixed | No |
| **Disease dynamics** | | | | | |
| | Transition rates (annual) | | | | |
| $l$ | Scale parameter of Weibull hazard | 2.86e-06 | (27) | Log-normal(m=-11.97, s=0.59) | Yes |
| $\gamma$ | Shape parameter of Weibull hazard | 2.78 | (27) | Log-normal(m=1.04, s=0.18) | Yes |
| $\lambda_2$ | Small adenoma to large adenoma | 0.0346 | (27) | Log-normal(m=-3.45, s=0.59) | Yes |
| $\lambda_3$ | Large adenoma to preclinical early CRC | 0.0215 | (27) | Log-normal(m=-3.91, s=0.35) | Yes |
| $\lambda_4$ | Preclinical early CRC to preclinical late CRC | 0.3697 | (27) | Log-normal(m=-1.15, s=0.23) | Yes |
| $\lambda_5$ | Preclinical early CRC to clinical late CRC | 0.2382 | (27) | Log-normal(m=-1.41, s=0.10) | Yes |
| $\lambda_6$ | Preclinical late CRC to clinical late CRC | 0.4582 | (27) | Log-normal(m=-0.78, s=0.22) | Yes |
| $\lambda_7$ | CRC mortality in early stage | 0.0302 | (27) | Fixed | No |
| $\lambda_8$ | CRC mortality in late stage | 0.2099 | (27) | Fixed | No |
| $\mu(a)$ | Age-specific mortality | Age-specific | (34) | Fixed | No |

Table 1. Description of parameters of the microsimulation model of the natural history of colorectal cancer.

*Calibration of the microsimulation model of the NH of CRC*

To state the calibration of the microsimulation model as an estimation problem (12), we define $M$ as the microsimulation model of the NH of CRC that has 11 input parameters $\theta \in \mathbb{R}^{11}$. Mortality rates from early and late stages of CRC could be obtained from cancer population registries (e.g., SEER in the U.S.), so there is no need to calibrate these. That is, $\theta_k = [\lambda_7, \lambda_8]$ is a set of 2 parameters that are known or could be obtained from external data (i.e., are external parameters). The model has a set of 9 parameters $\theta_u = [p_{adeno}, p_{small}, l, \gamma, \lambda_2, \lambda_3, \lambda_4, \lambda_5, \lambda_6]$ that cannot be directly estimated from sample data and need to be calibrated. The total number of parameters of $M$ is $\theta = [\theta_u, \theta_k]$.

To calibrate $M$, we adopted a Bayesian approach that allowed us to obtain a joint posterior distribution that characterizes the uncertainty of both the calibration targets and previous knowledge of the parameters of interest in the form of prior distributions. Prior distributions can reflect expert opinion or when little knowledge is available, these could be specified as uniform distributions. Following previous specifications, we constructed the likelihood function by assuming that the targets, $y_{t_i}$, are normally distributed with mean $\phi_{t_i}$ and standard deviation $\sigma_{t_i}$ (12). That is,

$$y_{t_i} \sim \text{Normal}(\phi_{t_i}, \sigma_{t_i}),$$

where $\phi_{t_i} = \mathbf{E}[M(\theta)]$ is the expected value of the model-predicted output for each type of target $t$ and age group $i$ at parameter set $\theta$. To compute an aggregated likelihood measure, we added the log-likelihoods across all targets. We defined prior distributions for all $\theta_u$ based on previous knowledge or nature of the parameters. For the prevalence of adenomas and proportion of small adenomas at age 50, we defined beta distributions, so they are bounded between 0 and 1. For the annual transition rates, which are defined over positive numbers, we assumed that their prior distributions follow a log-normal distribution. The prior distributions for these parameters are shown in Table 1. The ranges given in Table 1 are assumed to represent the 95% equal-tailed interval for the beta and log-normal distributions.

To conduct the Bayesian calibration, we used the incremental mixture importance sampling (IMIS) algorithm (35), which has been previously used to calibrate deterministic health policy models (17). An advantage of IMIS over other Monte Carlo methods, such as Markov chain Monte Carlo, is that with IMIS the evaluation of the likelihood for different sampled parameter sets could be parallelized, which makes its implementation perfectly suitable for an HPC environment using EMEWS.

*Propagation of uncertainty*

To characterize the joint uncertainty of the seven calibrated model parameters, we sampled 5,000 parameter sets from their joint posterior distribution using the IMIS algorithm. To compare the outputs of the calibrated model against the calibration targets, we propagated the uncertainty of the calibrated parameters through the microsimulation model of the NH of CRC by simulating a cohort of 100,000 individuals and computing the model-predicted adenoma and cancer outcomes for each of the 5,000 calibrated parameter sets drawn from their joint posterior distribution. To estimate the uncertainty limit model outputs, we computed the 95% posterior predicted interval (PI) defined as the estimated range between the $2.5^{th}$ and $97.5^{th}$ percentiles of the model-predicted posterior outputs.

*Cost-effectiveness analysis of screening for CRC*

With the calibrated microsimulation model of the NH of CRC, we assessed the cost-effectiveness of 10-yearly colonoscopy screening starting at age 50 years compared to no screening using a stochastic DM. For adenomas detected with colonoscopy, a polypectomy was performed during the procedure. Individuals diagnosed with a low- or high-risk polyp (i.e., small or large adenoma, respectively) underwent surveillance with colonoscopy every 5 or 3 years, respectively. We assumed screening or surveillance continued until 85 years of age. Recurrence rates after polypectomy were higher for individuals with a history of a polyp diagnosis. Only the transition rate from normal to small adenoma (i.e., $\lambda_1(a)$) was increased among those with a history of polyp. We assumed a hazard ratio (HR) of 2 for low-risk polyp and a HR of 3 for high-risk polyp. The costs of CRC treatment and utilities for CRC varied by stage. Patients without clinical CRC had a utility of 1. Table 2 shows the parameters used in the CEA with their corresponding distributions.

| Parameter | Value (Range) | Distribution | Source |
|---|---|---|---|
| **Screening test characteristics (location-specific)** | | | |
| **Small adenomas** | | | |
| Sensitivity | 0.773 (0.734-0.808) | Beta | (35) |
| Specificity | 0.868 (0.855-0.880) | Beta | (36) |
| **Large adenomas and CRC** | | | |
| Sensitivity | 0.950 (0.920-0.990) | Beta | (35) |
| Specificity | 0.868 (0.855-0.880) | Beta | (36) |
| **Increased rates after polypectomy (hazard ratio)** | | | |
| Low risk | 2 (1-3) | Log-normal | Assumed |
| High risk | 3 (2-4) | Log-normal | Assumed |

|  | Costs ($) | | |
| --- | --- | --- | --- |
| Colonoscopy | 10,000 (9,000-11,000) | Log-normal | Assumed |
| Early clinical CRC, annual costs | 21,524 (20,000-23,000) | Log-normal | Assumed |
| Late clinical CRC, annual costs | 37,000 (35,000-39,000) | Log-normal | Assumed |
|  | Utilities | | |
| Preclinical CRC | 1.000 (0.980-1.000) | Log-normal | Assumed |
| Early clinical CRC | 0.855 (0.700-0.900) | Log-normal | (37) |
| Late clinical CRC | 0.300 (0.200-0.400) | Log-normal | (37) |

Table 2. Description of cost-effectiveness analysis parameters

We conducted different PSA of the 10-year colonoscopy screening strategy for different approaches to characterize uncertainty of the two types of parameters, the calibrated and external (i.e., CEA parameters) using EMEWS to distribute the samples of each PSA across HPC resources. The first approach considers uncertainty in both types of parameters. The uncertainty of the calibrated parameters was characterized by their joint posterior distribution obtained from the IMIS algorithm. The second approach only considers uncertainty in the external parameters while fixing the calibrated parameters the *maximum-a-posteriori* (MAP) estimate (i.e., the parameter with the highest posterior density). The third approach considers uncertainty only in the calibrated parameters characterized by their joint posterior distribution and no uncertainty in the external parameters, which are fixed at their mean values. The fourth approach considers uncertainty in both types of parameters but compared to the first approach, the uncertainty of the calibrated parameters is characterized by constructing distributions based solely on posterior moments (i.e., means and standard deviations) and type of calibrated parameters ignoring correlations. We then estimated the EVPI of the 10-year colonoscopy screening strategy for the different approaches to characterize uncertainty of the calibrated and external parameters.

**RESULTS**

The IMIS algorithm was run on the Midway2 cluster at the University of Chicago Research Computing Center. Midway2 is a hybrid cluster, including both CPU and GPU resources. For this work, the CPU resources were used. Midway2 consists of 370 nodes of Intel E5-2680v4 processors, each with 28 cores and 64 GB of RAM. Using EMEWS we developed a workflow that parallelized the likelihood evaluations over 1008 processes using 36 compute nodes in total. We based the estimation of the calibrated parameters on 5,000 parameter sets sampled from the posterior distribution, including 3,241 unique parameter sets with an expected sample size (ESS) of 2,098.

Figure 2 shows the marginal prior and posterior distributions of the calibrated parameters. Table 3 gives estimated posterior means and standard deviations, MAP estimates and 95% credible intervals (CrI) for all calibrated parameters. From Table 3 and Figure 2, the prior and posterior means of the calibrate parameter do not seem that different but the major contrast is that the width of the posterior distributions shrunk, meaning that the calibration targets informed the calibrated parameters through a Bayesian updating.

| Parameter | Mean | SD | MAP | 95% CrI | |
|---|---|---|---|---|---|
| | | | | LB | UB |
| $p_{adeno}$ | 0.264 | 0.008 | 0.264 | 0.248 | 0.281 |
| $p_{small}$ | 0.706 | 0.019 | 0.711 | 0.667 | 0.741 |
| $l$ | 6.24E-06 | 3.16E-06 | 4.52E-06 | 1.92E-06 | 1.41E-05 |
| $\gamma$ | 2.639 | 0.112 | 2.635 | 2.432 | 2.877 |
| $\lambda_2$ | 0.035 | 0.002 | 0.035 | 0.031 | 0.039 |
| $\lambda_3$ | 0.021 | 0.001 | 0.021 | 0.020 | 0.023 |
| $\lambda_4$ | 0.374 | 0.036 | 0.368 | 0.310 | 0.448 |
| $\lambda_5$ | 0.247 | 0.021 | 0.251 | 0.209 | 0.288 |
| $\lambda_6$ | 0.457 | 0.076 | 0.435 | 0.345 | 0.664 |

Table 3. Posterior means, standard deviations, maximum-a-posteriori (MAP) estimate and 95% credible interval (CrI) of calibrated parameters of the microsimulation model of the natural history of colorectal cancer.

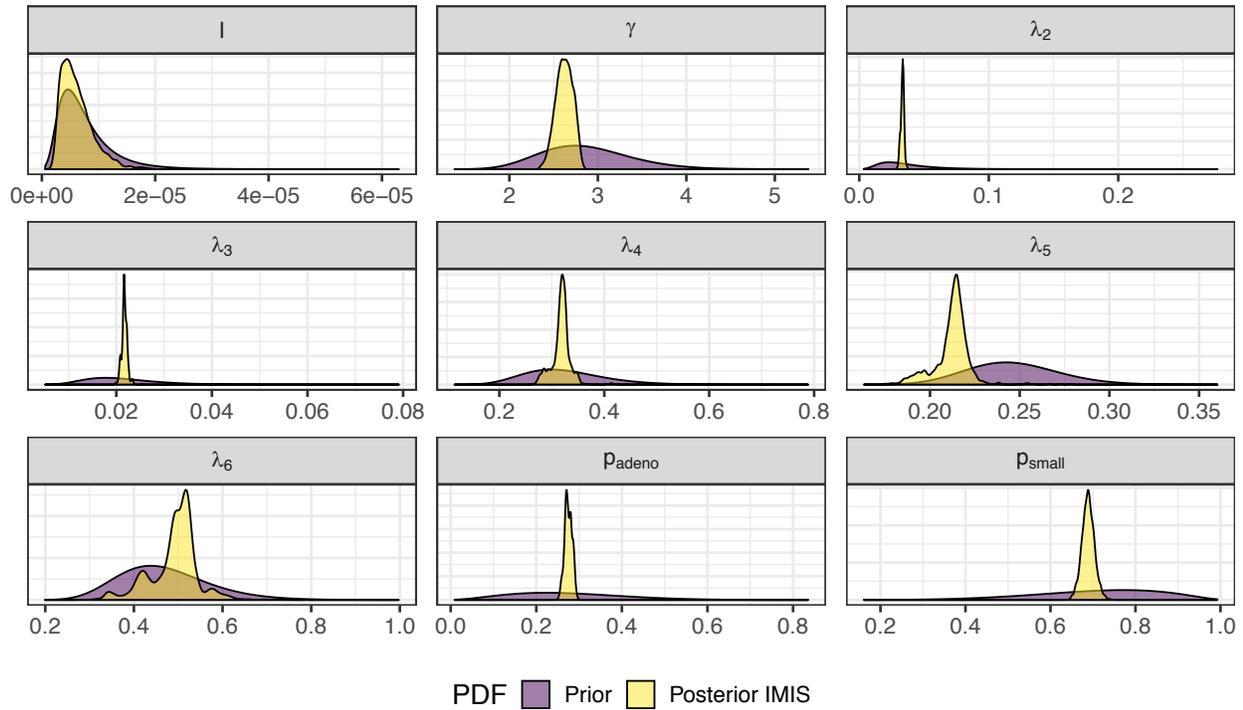

Figure 2. Prior and posterior marginal distributions of calibrated parameters of the microsimulation model of the natural history of colorectal cancer.

The Bayesian calibration also correlated the parameters showing the dependency among some of them, which can be seen on Figure 3 that shows the marginal and pair-correlations of the posterior distributions of the calibrated parameters. There are pairs of parameters with high correlation. The scale and shape parameters of the Weibull hazard function for the age of onset of adenomas, $l$ and $\gamma$, respectively, have the highest negative correlation of -0.958. The high correlation is the result from the calibration of the microsimulation model of the NH of CRC being non-identifiable when calibrating all 9 parameters to all the targets (see Alarid-Escudero et al. (12) for a detailed description of this analysis). The transition rates from preclinical early CRC to preclinical late and clinical early have a correlation of 0.784. The prevalence of adenomas and the proportion of small adenomas at age 50, which inform the initial distribution of the cohort across the adenoma health states, also have a high correlation of 0.482.

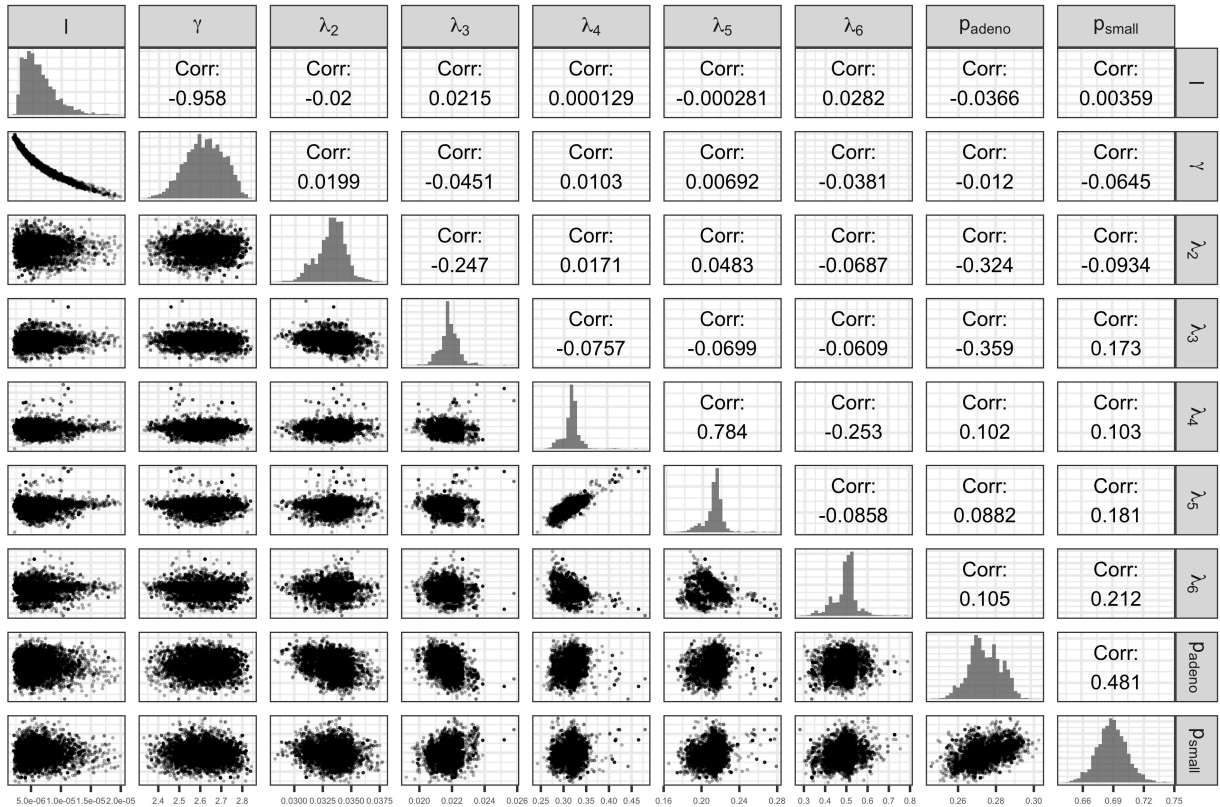

Figure 3. Scatter plot of pairs of calibrated parameters with correlation coefficient and posterior marginal distributions.

Figure 4 shows the internal validation of the calibrated model by comparing calibration targets with their 95% confidence interval (CI) and the model-predicted posterior means together with their 95% posterior PI. The calibrated model accurately predicted the calibration targets for both the means and the uncertainty intervals.

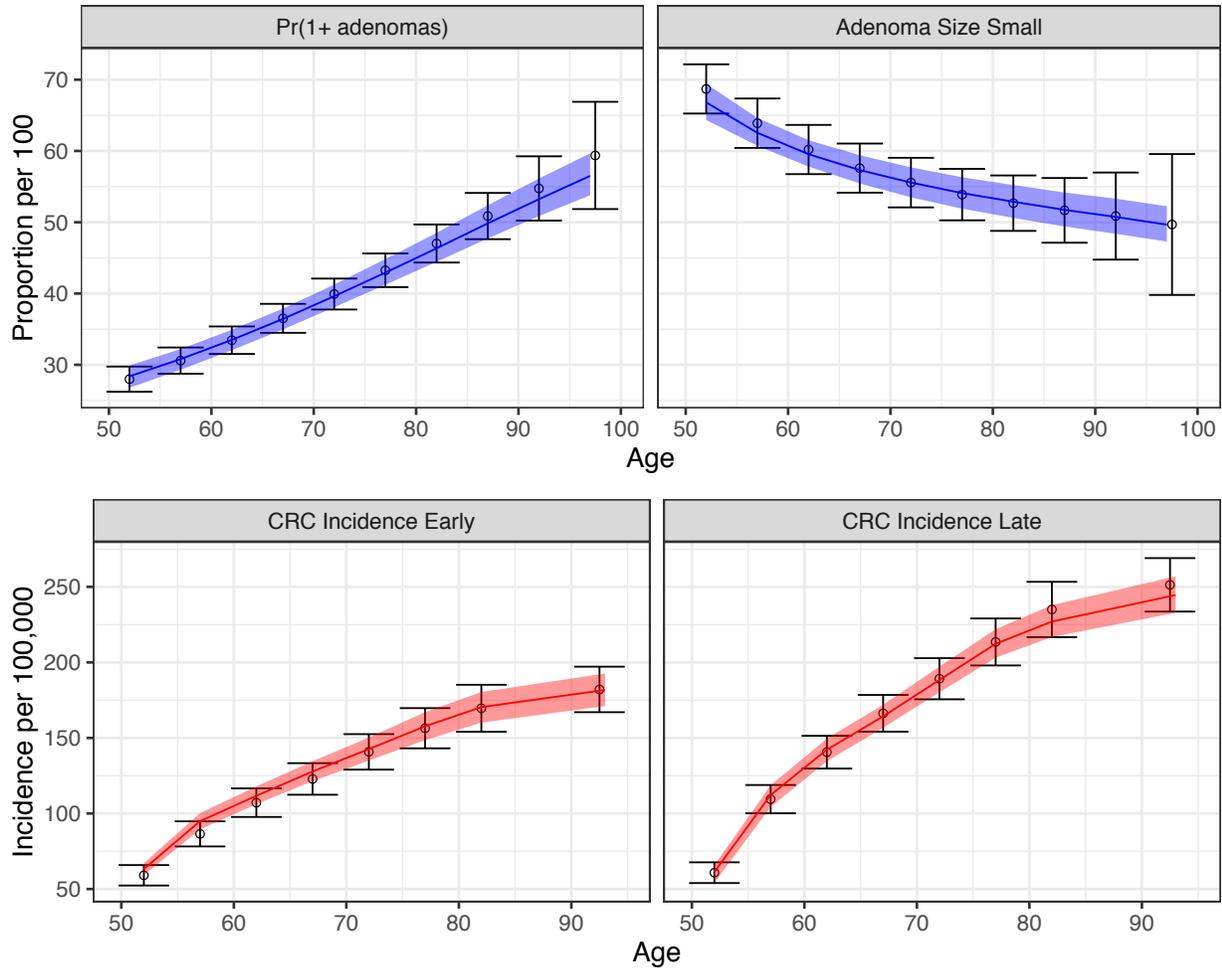

Figure 4. Comparison between posterior model-predicted outputs and calibration targets. Calibration targets with their 95% CI are shown in black. The shaded area shows the 95% posterior model-predictive interval of the outcomes and colored lines shows the posterior model-predicted mean based on 5,000 simulations using samples from posterior distribution. Upper panel refers to adenoma-related targets and lower panel refers to CRC incidence targets by stage.

The joint distribution of the incremental QALYs and incremental costs of the 10-year colonoscopy screening strategy resulting from the PSA for the different approaches to characterize uncertainty of the calibrated parameters are shown in Figure 5Figure *5* When accounting for uncertainty on the external parameters, there is little difference on the spread of the CEA outcomes between considering the joint distribution of the calibrated parameters and using only the MAP estimates (approaches 1 and 2 on the top row of Figure 5, respectively). The joint distribution of the outcomes is slightly wider when considering uncertainty on all parameters compared to when fixing the calibrated parameters at their MAP estimate. The third approach reflects the impact of only varying the calibrated parameters on the joint distribution of incremental QALYS and incremental costs, which is much more narrow compared to approaches 1 and 2. The fourth approach, which characterizes uncertainty of the calibrated parameters using method of moments without accounting for correlation, has the widest spread on the distribution of the outcomes.

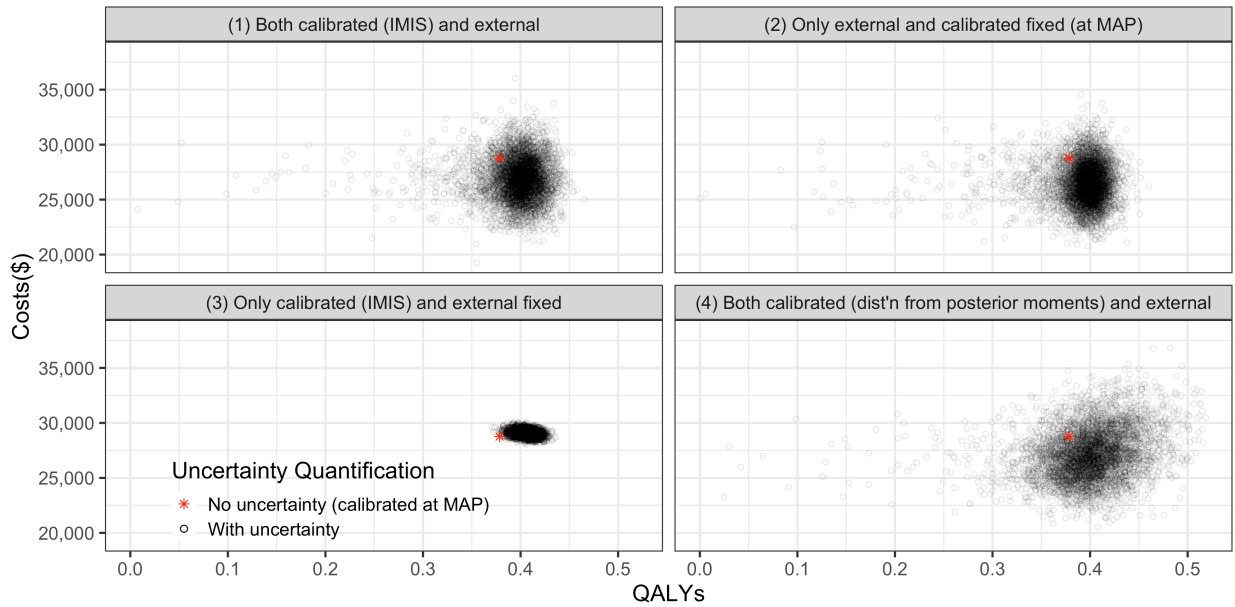

Figure 5. Incremental costs and incremental QALYs of 10-year colonoscopy screening vs no screening under different assumptions of characterization of uncertainty of both calibrated and external parameters. Red star corresponds to the incremental costs and incremental QALYs evaluated at the maximum-a-posteriori estimate of the calibrated parameters and the mean values of the external parameters.

Figure 6 shows the per-person EVPI of the 10-year colonoscopy screening strategy as a function of the willingness-to-pay (WTP) threshold for the four different approaches to characterize uncertainty of the calibrated and the external parameters. The first and second approach had similar EVPI up to reaching their maximum of $653 and $685, respectively, at a WTP threshold of $66,000/QALY. For WTP thresholds greater than $66,000/QALY, the first approach had higher EVPI than the second approach. When only uncertainty is considered for the calibrated parameters (approach 3), the EVPI is the lowest across all WTP thresholds with an EVPI of $0.1 at a WTP threshold of $66,000/QALY and reaching its highest of $212 at a WTP threshold of $71,000/QALY. The fourth approach reaches a maximum of $809 at a WTP threshold of $66,000/QALY and is the highest compared to the other approaches up to a WTP threshold of $81,000/QALY at which the first approach has the highest EVPI.

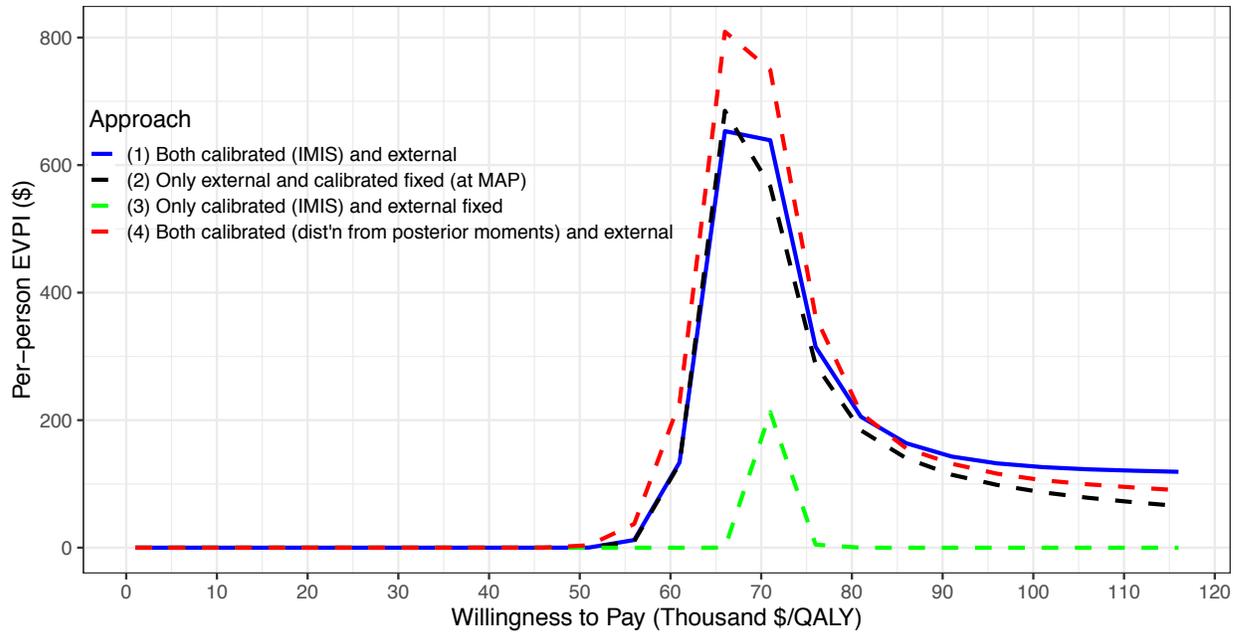

Figure 6. Per-patient EVPI of 10-year colonoscopy screening vs. no screening under different approaches to characterize uncertainty of both the calibrated and external parameters.

**DISCUSSION**

In this study, we characterized the uncertainty of a relatively simple but realistic microsimulation model of the NH of CRC by calibrating its parameters to different targets with varying degrees of uncertainty using a Bayesian approach on an HPC environment using EMEWS and quantified the value of the uncertainty of the calibrated parameters on the cost-effectiveness of a 10-year colonoscopy screening strategy. Although Bayesian calibration is a computationally intensive task, we took advantage of the parallelizability of the IMIS algorithm to efficiently conduct this task by evaluating the likelihood of different parameter sets in multiple cores simultaneously on an HPC setup. EMEWS has been previously used to calibrate other stochastic DMs (39,40) but has not been previously used to conduct PSA and calculate the VOI of calibrated parameters.

We found that different characterizations of uncertainty of calibrated parameters have implications on the expect value of reducing uncertainty on the CEA. Ignoring inherent correlation among calibrated parameters on a PSA overestimates the value of uncertainty. When full posterior distributions of the calibrated parameters are not readily available, the MAP could be considered as a best parameter set. In our example, not considering uncertainty of calibrated parameters on the PSA did not seem to have a meaningful impact on the uncertainty of the CEA outcomes and the EVPI of the screening strategy. That is, the uncertainty associated with the NH was less valuable than the uncertainty of the external parameters. However, these results should be taken with caution because this analysis is conducted on a fictitious model with simulated calibrated targets. The impact of a well conducted characterization of the uncertainty of calibrated parameters on CEA outcomes and VOI measures should be analyzed on a case-by-case basis.

There are examples of calibrated parameters being included in a PSA. For example, by taking a certain number of good-fitting parameter sets (41,42), bootstrapping with equal probability good-fitting parameter sets obtained through directed search algorithms (e.g., Nelder-Mead) (43), or conducting a Bayesian calibration, which produces the joint posterior distribution of the calibrated parameters (17).However, this is the first manuscript to conduct a PSA and VOI analysis using distributions of calibrated parameters of a stochastic DM that accurately characterize their uncertainty.

Currently, Bayesian calibration of stochastic DMs might not be feasible on regular desktop or laptop computers. To circumvent current computational limitations from using Bayesian methods in calibrating stochastic DM, the use of surrogate models -often called metamodels or emulators- has been proposed (44,45). Surrogate models are statistical models like Gaussian processes (46) or neural networks (47) that aim to replace the relation between inputs and outputs of the original stochastic DM (48,49), which once fitted they are computationally more efficient to run than the stochastic DM. Constructing an emulator might not be a straightforward task because the stochastic DM still needs to be evaluated at different parameter sets which could also be computationally expensive. Furthermore, the statistical routines to build the emulator may not be readily available in the programming language in which the stochastic DM is coded. These are the type of situation where EMEWS can be used to efficiently construct metamodels. This, however, is a topic for further research

In this article, we showed that EMEWS can facilitate the use of HPC to implement computationally demanding Bayesian calibration routines to correctly characterize the uncertainty of the calibrated parameters of stochastic DMs and propagate it in the evaluation of CEA of screening strategies and quantify their value of information. The methodology and results of this study could be used to guide a similar VOI analysis on a more realistic, hence more complicated, stochastic DM of CRC screening to determine where more research is needed and guide research prioritization.

## ACKNOWLEDGEMENTS

Financial support for this study was provided in part by a grant from the National Council of Science and Technology of Mexico (CONACYT) and a Doctoral Dissertation Fellowship form the Graduate School of the University of Minnesota as part of Dr. Alarid-Escudero's doctoral program. Drs. Kuntz, Knudsen and Alarid-Escudero were supported by a grant from the National Cancer Institute (U01-CA-199335) as part of the Cancer Intervention and Surveillance Modeling Network (CISNET). The funding agencies had no role in the design of the study, interpretation of results, or writing of the manuscript. The content is solely the responsibility of the authors and does not necessarily represent the official views of the National Institutes of Health. The funding agreement ensured the authors' independence in designing the study, interpreting the data, writing, and publishing the report.

## REFRENCES